\documentclass{article}

\usepackage[final]{neurips_2024}

\usepackage[utf8]{inputenc} 
\usepackage[T1]{fontenc}    
\usepackage{hyperref}       
\usepackage{url}            
\usepackage{booktabs}       
\usepackage{amsfonts}       
\usepackage{nicefrac}       
\usepackage{microtype}      
\usepackage{xcolor}         
\usepackage{graphicx}
\usepackage{makecell}  
\usepackage{multirow}
\usepackage{arydshln}
\usepackage{booktabs}       
\usepackage{color}
\usepackage{caption}
\usepackage{float}
\usepackage{amsfonts}
\usepackage{hyperref}
\usepackage[htt]{hyphenat}
\usepackage{enumitem}
\usepackage{array}     
\usepackage{float}

\newcommand{\ourmethod}{\textit{Pixel Perfect MegaMed}}

\usepackage{pifont}
\newcommand{\cmark}{\ding{51}}  
\newcommand{\xmark}{\ding{55}}  

\author{%
  Zahra ~TehraniNasab\\
  McGill University\\
  MILA-Quebec AI Institute\\
  \texttt{zahra.tehraninasab@mail.mcgill.ca} \\
  \And
  Hujun ~Ni\\
  McGill University\\
  \texttt{hujun.ni@mail.mcgill.ca} \\
  \And
  Amar ~Kumar\\
  McGill University\\
  MILA-Quebec AI Institute\\
  \texttt{amar.kumar@mail.mcgill.ca} \\
  \And
  Tal ~Arbel\\
  McGill University\\
  MILA-Quebec AI Institute\\
  \texttt{tal.arbel@mcgill.ca} \\
}

\begin{document}
%
\title{Pixel Perfect MegaMed: A Megapixel-Scale Vision-Language Foundation Model for Generating High Resolution Medical Images} 

\maketitle              
\begin{abstract}
Medical image synthesis presents unique challenges due to the inherent complexity and high-resolution details required in clinical contexts. Traditional generative architectures such as Generative Adversarial Networks (GANs) or Variational Auto Encoder (VAEs) have shown great promise for high-resolution image generation but struggle with preserving fine-grained details that are key for accurate diagnosis. To address this issue, we introduce \ourmethod, the first vision-language foundation model to synthesize images at resolutions of $1024 \times 1024$. Our method deploys a multi-scale transformer architecture designed specifically for ultra-high resolution medical image generation, enabling the preservation of both global anatomical context and local image-level details. By leveraging vision-language alignment techniques tailored to medical terminology and imaging modalities, \ourmethod\ bridges the gap between textual descriptions and visual representations at unprecedented resolution levels. We apply our model to the CheXpert dataset and demonstrate its ability to generate clinically faithful chest X-rays from text prompts. Beyond visual quality, these high-resolution synthetic images prove valuable for downstream tasks such as classification, showing measurable performance gains when used for data augmentation, particularly in low-data regimes. Our code is accessible through the project website\footnote{https://tehraninasab.github.io/pixelperfect-megamed/ \\}.

\end{abstract}


%
%
%
\section{Introduction}
\begin{figure}[t]
    \centering
    \includegraphics[width=0.9\linewidth]{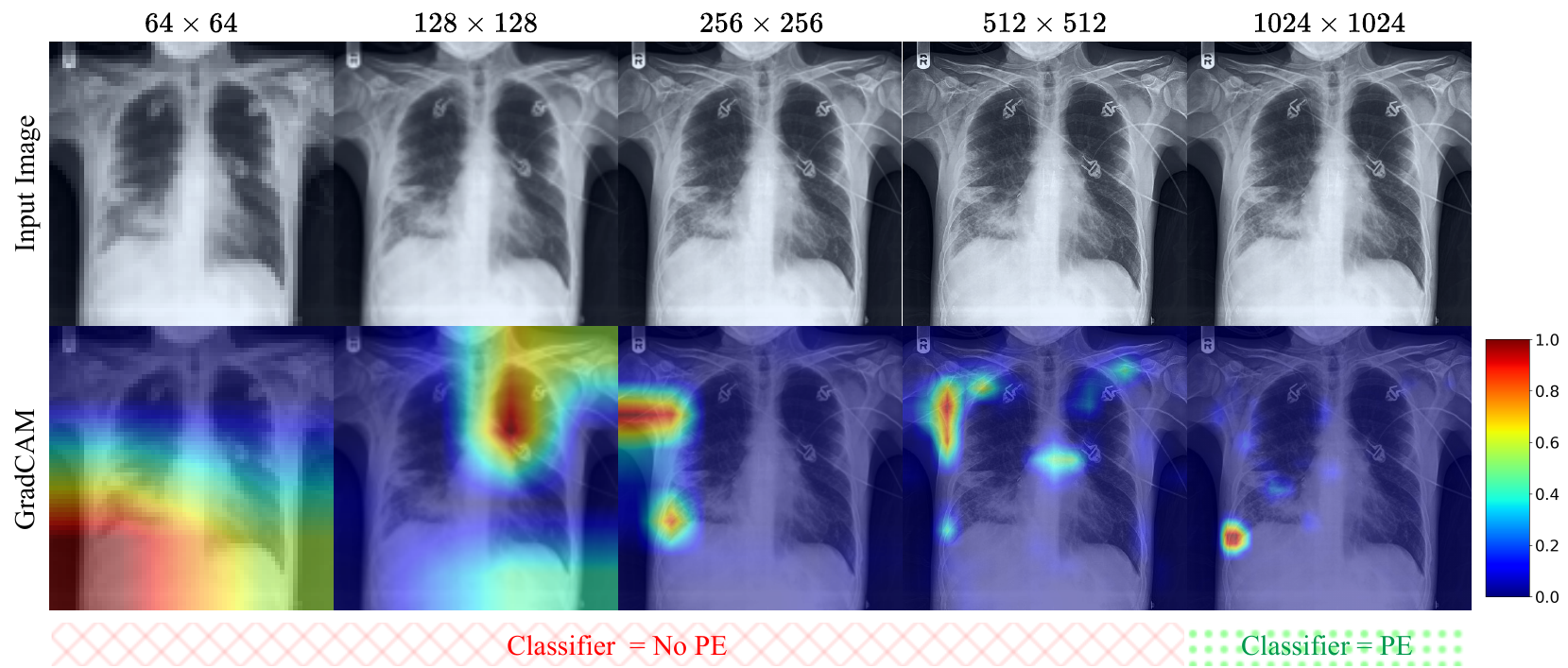}
    \caption{Grad-CAM visualization of activation maps in the EfficientNet classifier for pleural effusion (PE) classification of a patient's Chest X-ray image~\cite{irvin2019chexpert}. The heatmaps highlight that the model fails to focus on relevant regions in low-resolution images, leading to incorrect classifications. At the higher resolution, the model focuses on the exact location of interest.}
    \label{fig:introduction}
\end{figure}

\begin{figure}[t]
    \centering
    \includegraphics[width=0.85\linewidth]{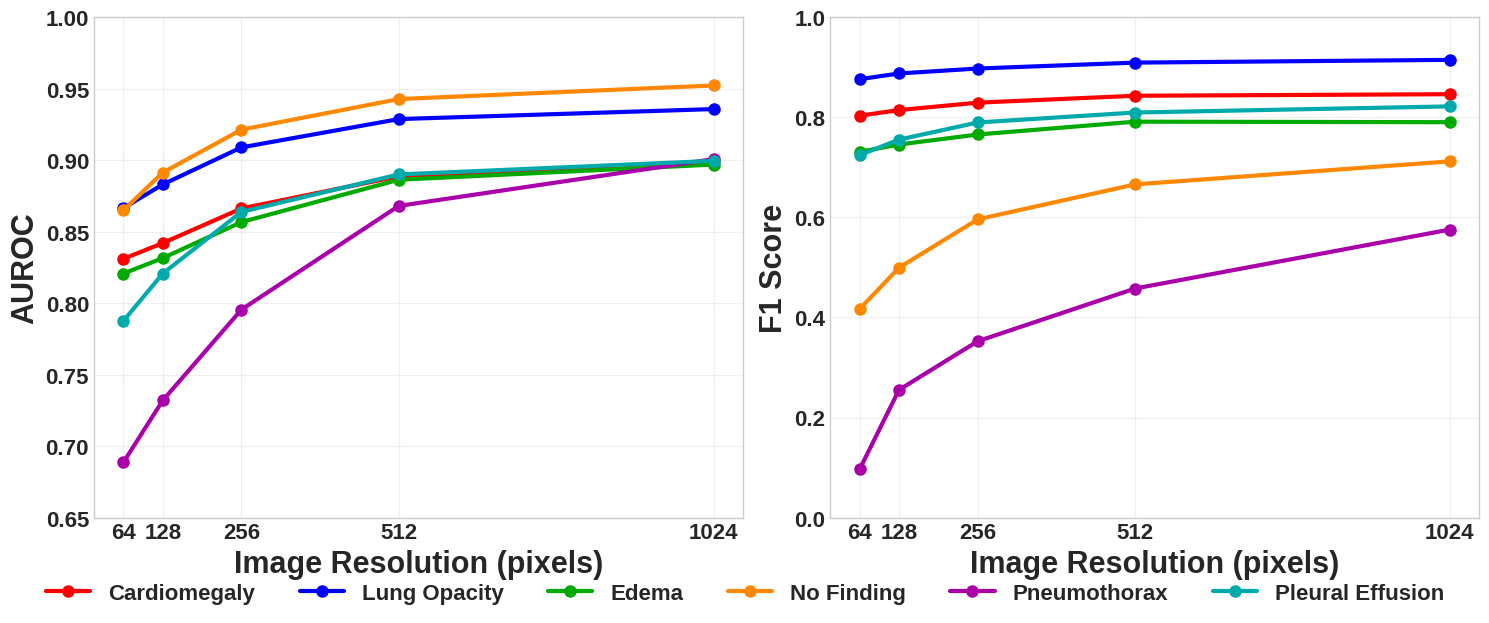}
    \caption{The effect of image resolution on multi-class disease classification performance (Left: AUROC, Right: F1). The same samples across different resolutions are used to train all the EfficientNet classifiers.}
    \label{fig:motivation}
\end{figure}

High-resolution medical images are needed for many clinical decision support systems that depend on the ability to resolve fine-grained anatomical and pathological features. Consider its importance in the context of chest X-rays, where subtle abnormalities—such as small pulmonary nodules, fine patterns, or early pleural changes—are more easily identified at higher resolutions~\cite{haque2023effect,jiang2025high}. AI-driven diagnostic systems should maintain images at high resolution, if available, in order to preserve essential texture and edge information that might otherwise be lost at lower resolutions~\cite{miyata2020influence,schuijf2022ct,yanagawa2018subjective}. 
For instance, in detecting pleural effusion, the separation of the pleural line—a key diagnostic indicator—may remain undetectable at lower resolutions but becomes visible when sufficient spatial detail is present (see Figure~\ref{fig:introduction} \& \ref{fig:motivation}). Given their importance, high-resolution image generators can synthesize detailed medical images in contexts where real, high-quality data is scarce, thereby supporting the development of robust diagnostic models in data-limited settings.

Image-based generative models have made significant advances in medical imaging, holding substantial potential for advancing analysis and enabling data augmentation for improved classification~\cite{fathi2024decodex,kumar2023debiasing} and segmentation~\cite{chlap2021review,chen2022enhancing}. Conditional generative models have led to huge advances in explainability through counterfactual image generation~\cite{fathi2024decodex,mertes2022ganterfactual}, leading to advances in understanding personalized markers of disease~\cite{kumar2022counterfactual}. Furthermore, recent advances in VLM foundation models (e.g. Stable Diffusion) have permitted significant performance improvements for many tasks when fine-tuned on medical images. 
However, most existing work in medical image synthesis has been constrained to low or moderate resolutions, typically around $128 \times 128$~\cite{iklima2022realistic,madani2018chest} or $256 \times 256$ pixels~\cite{atad2022chexplaining,fathi2024decodex}. These resolutions are inadequate for clinical use, as they fail to capture the detailed anatomical structures and subtle pathological cues necessary for accurate diagnosis (see Figure~\ref{fig:motivation}). Although recent methods have started to push resolution boundaries—reaching $512 \times 512$ pixels~\cite{kumar2025prism,perez2025radedit}—generating high-quality, high-resolution medical images that retain clinical utility remains a significant challenge. Achieving ultra-high resolution synthesis (e.g., $1024 \times 1024$ and beyond) is crucial for capturing the full complexity of medical imagery, including tiny anatomical variations and rare pathological signatures that are vital for robust diagnostic and research applications.

In this work, we present \ourmethod, the first VLM foundation model capable of synthesizing ultra-high-resolution medical images at $1024 \times 1024$ pixels, setting a new benchmark (4 times larger than existing VLM) for fidelity and clinical relevance in generative medical imaging. Our approach builds upon a multi-scale transformer-based backbone architecture based on Stable-Diffusion XL (SDXL)~\cite{podell2023sdxl}, a VLM for synthesizing ultra-high resolution images, fine-tuned on a medical imaging dataset, CheXpert~\cite{irvin2019chexpert}. Furthermore, we extend our framework to progressively upscale the generated images to a resolution of $2048 \times 2048$, further pushing the boundaries of photorealism and clinical relevance. We validate our model's performance using rigorous image quality metrics, including Fréchet Inception Distance (FID) and Vendi Score, demonstrating synthesis fidelity and perceptual quality of our method. Additionally, we demonstrate that images generated by \ourmethod\ can be used for data augmentation, yielding measurable gains in downstream classification performance under limited data regimes. The code and model weights of our new VLM models will be released to permit widespread adoption in the medical imaging community. 



\section{Methodology}
\begin{figure}[t]
    \centering
    \includegraphics[width=0.9\linewidth]{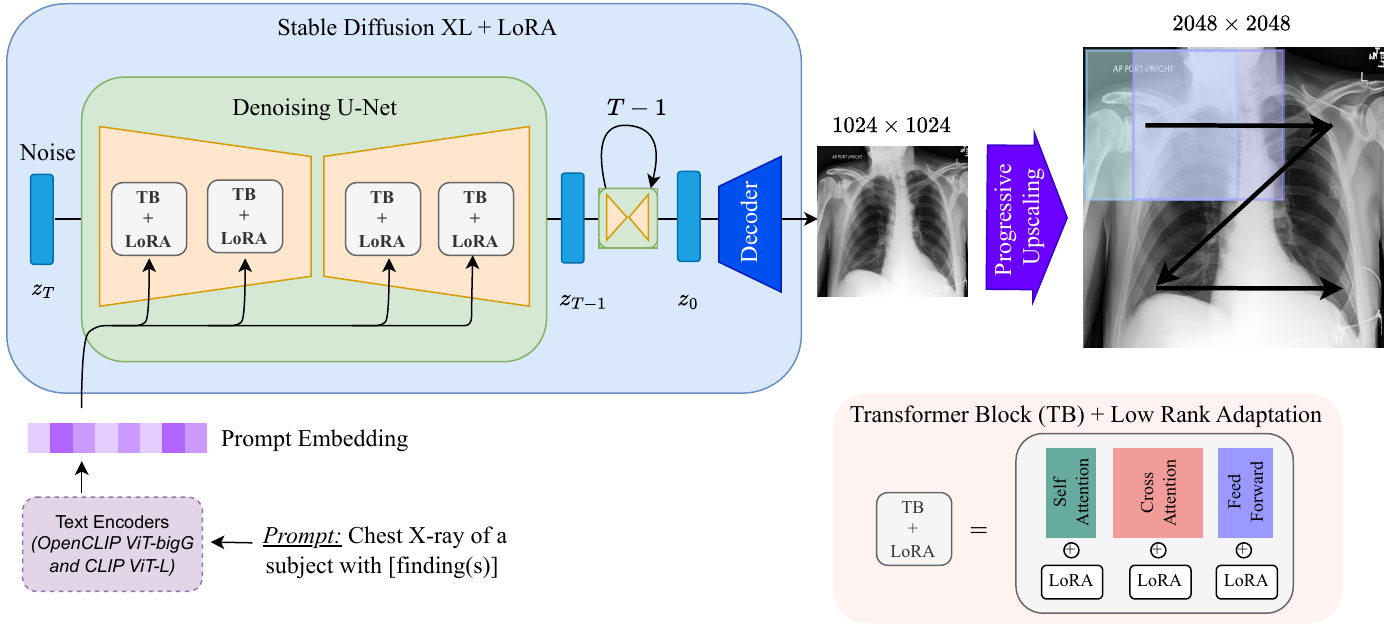}
    \caption{Architecture for high-resolution medical image synthesis using SDXL~\cite{podell2023sdxl}.}
    \label{fig:arch}
\end{figure}

In this work, we propose building a VLM foundation model for high-resolution medical images by fine-tuning SDXL using Low-Rank Adaptation (LoRA)~\cite{hu2022lora}. Our framework includes: (i) generation of a $1024\times1024$ image conditioned on embeddings from OpenCLIP ViT-bigG~\cite{ilharco10openclip} and CLIP ViT-L~\cite{radford2021learning}; (ii) refinement via a denoising module to improve anatomical realism; (iii) [\textit{optional}] a progressive upscaling module increases the resolution to $2048\times2048$, see Figure~\ref{fig:arch}.

\subsection{MultiDiffusion: Adapted Latent Diffusion}
Latent Diffusion Models (LDMs) ~\cite{Rombach_2022_CVPR} perform diffusion in a learned latent space rather than directly in the high-dimensional pixel space, thus reducing the computational costs while preserving the quality of generated samples. Given a pre-trained autoencoder with an encoder $\mathcal{E}$ and decoder $\mathcal{D}$, images $\mathbf{x} \in \mathbb{R}^{H \times W \times C}$ are first mapped into a lower-dimensional latent space as $\mathbf{z} = \mathcal{E}(\mathbf{x})$, $\mathbf{z} \in \mathbb{R}^{h \times w \times c}$. A standard diffusion process is then applied in the latent space to model the data distribution. The forward diffusion process gradually adds Gaussian noise to the latent variable $\mathbf{z}_0$ over $T$ steps: $
    q(\mathbf{z}_t \mid \mathbf{z}_{t-1}) = \mathcal{N}(\mathbf{z}_t; \sqrt{1 - \beta_t} \mathbf{z}_{t-1}, \beta_t \mathbf{I}),
$

\noindent where $\beta_t$ is a fixed variance schedule. The model learns a denoising function $\epsilon_\theta(\mathbf{z}_t, t)$ to approximate the added noise, enabling the reverse process to recover $\mathbf{z}_0$ iteratively. The final image is reconstructed via $\hat{\mathbf{x}} = \mathcal{D}(\mathbf{z}_0)$. 

\textbf{MultiDiffusion}~\cite{bar2023multidiffusion} enhances latent diffusion models (LDMs) by enabling the image generation at resolutions higher than those originally trained on. Rather than synthesizing the entire image in a single step, the method breaks the target canvas into several overlapping tiles. Each tile is processed independently in the latent space, guided by a shared prompt or contextual information. Once individual tiles are denoised, they are combined using a weighted averaging technique that smooths the overlaps and ensures visual consistency. This tiling and merging strategy enables the model to maintain both fine-grained local details and a coherent global structure, allowing for high-resolution image synthesis without the need to retrain the underlying diffusion model.
\subsection{Finetuning SDXL}
A pre-trained latent diffusion model, SDXL~\cite{podell2023sdxl}, was finetuned using LoRA~\cite{hu2022lora} to adapt the model to domain-specific image generation tasks efficiently. LoRA introduces trainable low-rank matrices into the attention layers of the U-Net (and optionally the text encoder), enabling parameter-efficient fine-tuning without modifying the original weights. Formally, for a given attention layer with query/key/value projection matrices $M \in \mathbb{R}^{d \times d}$, LoRA adds a low-rank update in the form $M' = M + \Delta M$, where $\Delta M = A B$ with $A \in \mathbb{R}^{d \times r}$ and $B \in \mathbb{R}^{r \times d}$, and $r \ll d$. Only $A$ and $B$ are optimized during fine-tuning, while the original weights, $M$, remain frozen, resulting in a significant reduction in trainable parameters from $\mathcal{O}(d^2)$ to $\mathcal{O}(2dr)$.

LoRA enables learning domain-specific concepts such as medical conditions (e.g., "Cardiomegaly", "Pneumothorax") and aligning the model's output with semantic prompts. Similar to~\cite{kumar2025prism}, the binary labels from the CheXpert dataset are converted into textual prompts to fine-tune SDXL~\cite{podell2023sdxl}. Conditioning is performed using concatenated embeddings from OpenCLIP ViT-bigG~\cite{ilharco10openclip} and CLIP ViT-L~\cite{radford2021learning}, with training prompts as: \texttt{Chest X-ray of a subject with [finding(s)]}. The findings include: \textit{No Finding, Enlarged Cardiomediastinum, Cardiomegaly, Lung Opacity, Lung Lesion, Edema, Consolidation, Pneumonia, Atelectasis, Pneumothorax, Pleural Effusion, Pleural Other, Fracture, and Support Devices}.

As an additional feature, progressive upscaling module can be used to further enhance the resolution of the generated images from $1024\times1024$ to $2048\times2048$, similar to the DemoFusion framework~\cite{du2024demofusion}. Instead of using a traditional one-shot super-resolution approach, we leverage an `upsample–diffuse–denoise' loop, where each phase progressively refines the image by introducing noise into an upsampled latent representation and denoising it through a pre-trained diffusion model. The skip residuals in the implementation provide global structural guidance, while dilated sampling ensures semantic coherence across local patches. 
\subsection{Evaluating Synthesized Ultra-High Resolution Images}
To assess the quality and utility of the synthesized ultra-high resolution medical images, we conduct both perceptual and downstream task-based evaluations.
\noindent \textbf{Perceptual Quality Metrics}: We employ standard evaluation metrics including the Fr\'echet Inception Distance (FID)~\cite{heusel2017gans} and Vendi Score (VS)~\cite{friedman2022vendi} to quantitatively assess the visual fidelity and diversity of the generated images. 
For FID, feature representations are extracted using the 1024-dimensional penultimate layer of the pre-trained DenseNet-121 model from the TorchXRayVision~\cite{cohen2022torchxrayvision} library, trained on a wide range of chest X-ray datasets.

\noindent \textbf{Downstream Classification Performance}: To assess the utility of synthetic images in clinical applications, we evaluate their effectiveness in enhancing classification performance through dataset augmentation under limited data conditions. For each target pathology, we sample 100 real images from the CheXpert dataset and augment them with 2,000 high-resolution ($1024 \times 1024$) synthesized images. A multi-label EfficientNet~\cite{tan2019efficientnet} classifier with six output heads is trained on this augmented dataset and evaluated on a held-out CheXpert dataset. An analogous augmentation strategy—using the same set of synthesized images generated from CheXpert—is applied to the MIMIC-CXR dataset, which contains the same set of target pathologies. This allows us to assess the generalizability of synthetic data across datasets and examine its impact on model performance under domain shift, where the training and evaluation distributions differ. 

\section{Experiments and Results}
\subsection{Dataset and Implementation Details}
We perform experiments on a publicly available dataset, CheXpert~\cite{irvin2019chexpert}, with a training/ validation/ test split of 70/ 15/ 15, see Table~\ref{table:data-distribution}. Additionally, we use the MIMIC-CXR~\cite{johnson2019mimic} dataset to evaluate the performance of augmented classifiers in a data-scarce scenario. Noise scheduling is performed using the Euler discrete scheduler during both training and inference. For fine-tuning, we apply LoRA modules to the self-attention, cross attention and feed-forward layers of the U-Net's transformer blocks. 
SNR-weighted loss~\cite{hang2023efficient} is employed during training with $\gamma = 5.0$, reweighting the training objective based on the signal-to-noise ratio. To support reproducibility and future research, the source code and model weights will be made publicly available.

\begin{table}[t]
\centering
\caption{Summary of the train, validation and test splits. Note: Individual images can reflect the presence of several concurrent diseases. For data augmentation experiments, testing is conducted on both datasets using the CheXpert augmented training set.}
\begin{tabular}{clclcccc}
\cline{3-3} \cline{5-8}
\multicolumn{1}{l}{} &  & \multicolumn{1}{l}{Finetuning} &  & \multicolumn{4}{c}{Classification} \\ \cline{3-3} \cline{5-8} 
 &  & CheXpert &  & \multicolumn{3}{c}{CheXpert} & MIMIC-CXR \\ \cline{1-1} \cline{3-3} \cline{5-8} 
\textbf{Class} &  & \textbf{Training} &  & \textbf{\begin{tabular}[c]{@{}c@{}}Training\\ Real + Synth\end{tabular}} & \textbf{Validation} & \textbf{Test} & \textbf{Test} \\ \cline{1-1} \cline{3-3} \cline{5-8} 
Cardiomegaly &  & 78149 &  & 100 + 2000 & 16682 & 16606 & 20490 \\
Lung Opacity &  & 96212 &  & 100 + 2000 & 20519 & 20608 & 22447 \\
Edema &  & 62036 &  & 100 + 2000 & 13251 & 13292 & 14457 \\
No Finding &  & 17014 &  & 100 + 2000 & 3644 & 3737 & 5131 \\
Pneumothorax &  & 15868 &  & 100 + 2000 & 3416 & 3519 & 6141 \\
Pleural Effusion &  & 65391 &  & 100 + 2000 & 13841 & 13956 & 28717 \\ \cline{1-1} \cline{3-3} \cline{5-8} 
\end{tabular}
\label{table:data-distribution}
\end{table}

\subsection{Results}
\noindent\textbf{Qualitative Evaluations} We present qualitative results of our method under two scenarios: (i) ultra-high resolution generation, showcasing the model’s ability to synthesize anatomically coherent and visually detailed medical images at $1024\times1024$ resolutions; and (ii) progressive upscaling, illustrating the refinement of semantic and structural details across successive resolution stages of $2048\times2048$ pixels. 
\begin{figure}[t]
    \centering
    \includegraphics[width=0.85\linewidth]{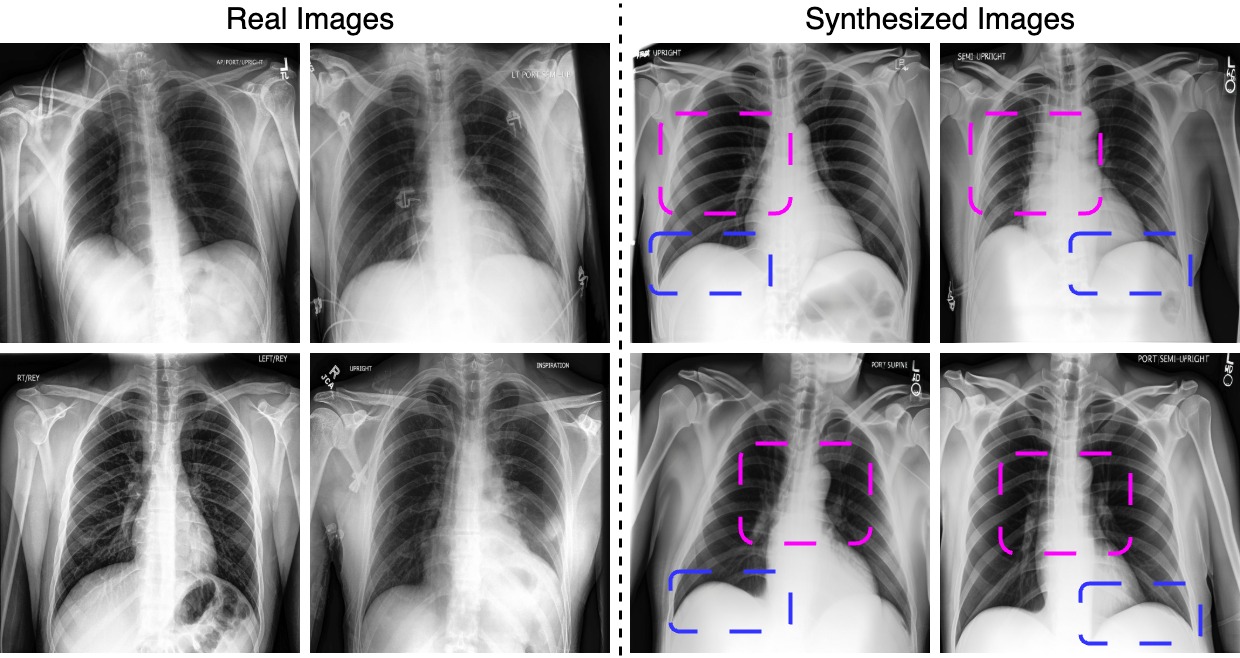}
    \caption{Comparison of (left) real samples and (right)  synthesized samples at $1024\times1024$ resolution. Note the preservation of fine-grained anatomical details such as subtle \textcolor[HTML]{FF00FF}{texture variations} in the lungs and sharp boundaries between  \textcolor[HTML]{3333FF}{anatomical regions}—features that are often lost or blurred at lower resolutions.}
    \label{fig:synthesized_samples}
\end{figure}
\begin{figure}
    \centering
    \includegraphics[width=0.95\linewidth]{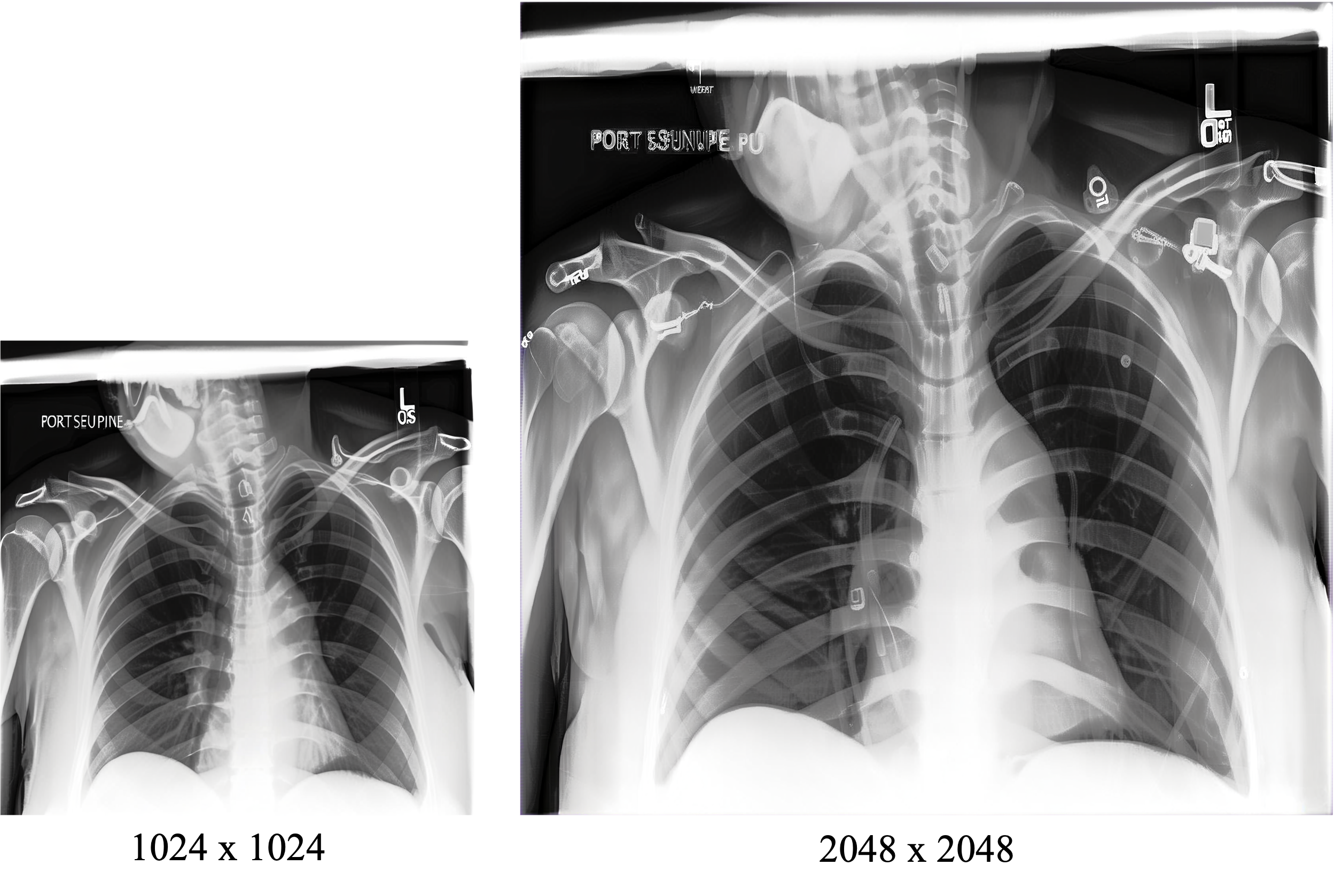}
    \caption{Progressively scaling an image from $1024\times1024$ to $2048\times2048$.}
    \label{fig:progressive-scaling}
\end{figure}

\noindent\textbf{Quantitative Evaluations} Table~\ref{tab:perclass_metrics} shows the image generation quality of high-resolution images synthesized using our technique. The FID scores are computed using 12,000 high-resolution synthetic samples (2,000 per class) along with the CheXpert test samples mentioned in Table~\ref{tab:perclass_metrics}. These scores can serve as quantitative benchmarks for future research in high-fidelity medical image generation. 
Table~\ref{table:augmentation} shows the impact of augmenting real clinical datasets with high-resolution synthetic images on classification performance across six disease categories in both CheXpert and MIMIC-CXR. Augmenting each class with 2,000 synthesized samples consistently improves both AUC-ROC and F1 scores across most categories. In CheXpert, the most notable improvement in F1 score is observed for Edema with gains of +0.054. Notably, several classes in MIMIC-CXR, such as Lung Opacity, Edema, No Finding and Pneumothorax, exhibit near-zero or zero F1 scores in the absence of augmentation, indicating the classifier’s inability to detect these pathologies under limited-data conditions. The introduction of synthetic data significantly mitigates this issue, resulting in F1 scores of 0.194, 0.336, 0.381 and 0.137, respectively. These findings suggest that high-resolution synthetic data not only enriches limited training sets but also enhances the model’s sensitivity to nuanced pathologies, thereby improving generalization to real-world clinical distributions.


\begin{table}[h]
\centering
\caption{Evaluation of image synthesis quality using FID and Vendi Score.}
\begin{tabular}{lcccccc}
\hline
\textbf{Metric}
  & \makecell{\textbf{Cardiomegaly}}
  & \makecell{\textbf{Lung}\\\textbf{Opacity}}
  & \textbf{Edema}
  & \makecell{\textbf{No}\\\textbf{Finding}}
  & \textbf{Pneumothorax}
  & \makecell{\textbf{Pleural}\\\textbf{Effusion}} \\
\hline
FID $\downarrow$       & 13.01 & 13.57 & 13.02 & 6.61  & 10.22 & 14.17 \\
Vendi Score $\uparrow$ &  3.08 &  2.83 &  2.89 & 2.80  &  2.98 &  3.10 \\
\hline
\end{tabular}
\label{tab:perclass_metrics}
\end{table}


\begin{table}[!htbp]
\centering
\caption{Performance of pretrained Efficient-Net~\cite{tan2019efficientnet} on a held-out test set after augmenting (100 real samples per class) with 2000 samples of $1024\times1024$ resolution synthetic samples (from CheXpert) per class.}
\begin{tabular}{cccclcc}
\hline
 &  & \multicolumn{2}{c}{CheXpert} &  & \multicolumn{2}{c}{MIMIC-CXR} \\ \hline
 & Augmentation & AUC-ROC & F1 &  & AUC-ROC & F1 \\ \cline{3-4} \cline{6-7} 
\multirow{2}{*}{Cardiomegaly} & \xmark & 0.814 & 0.794 &  & 0.583 & 0.331 \\
 & \cmark  & \textbf{0.831} & \textbf{0.807} &  & \textbf{0.619} & \textbf{0.409} \\ \hdashline
\multirow{2}{*}{Lung Opacity} & \xmark & 0.864 & 0.880 &  & 0.515 & 0.091 \\
 & \cmark  & \textbf{0.879} & \textbf{0.882} &  & \textbf{0.568} & \textbf{0.194} \\ \hdashline
\multirow{2}{*}{Edema} & \xmark & 0.817 & 0.701 &  & 0.620 & 0.052 \\
 & \cmark  & \textbf{0.842} & \textbf{0.755 }&  & \textbf{0.672} & \textbf{0.336} \\  \hdashline
\multirow{2}{*}{No Finding} & \xmark & 0.901 & 0.588 &  & 0.624 & 0.014 \\
 & \cmark  & \textbf{0.913} & \textbf{0.611} &  & \textbf{0.664} & \textbf{0.381} \\  \hdashline
\multirow{2}{*}{Pneumothorax} & \xmark & 0.703 & 0.307 &  & 0.587 & 0.020 \\
 & \cmark  & \textbf{0.742} & \textbf{0.308} &  & \textbf{0.656} & \textbf{0.137} \\  \hdashline
\multirow{2}{*}{Pleural Effusion} & \xmark & 0.766 & 0.666 &  & 0.609 & 0.511 \\
 & \cmark  & \textbf{0.785} & \textbf{0.680} &  & \textbf{0.620} & \textbf{0.595} \\ \hline
\end{tabular}
\label{table:augmentation}
\end{table}


\section{Conclusion}
In this work, we presented a framework for synthesizing ultra-high resolution medical images by fine-tuning SDXL using low-rank adaptation (LoRA) and incorporating a progressive upscaling module. By optimizing only a lightweight set of parameters, our approach efficiently learns clinically meaningful concepts from textual prompts while preserving the expressive power of large-scale pre-trained models. The integration of progressive upscaling—via an iterative `upsample–diffuse–denoise' process, skip residuals, and dilated sampling—enables the generation of anatomically coherent images at high resolutions. Through both quantitative metrics and downstream classification tasks, we demonstrate that the synthesized images not only exhibit high perceptual quality but also serve as valuable assets for data augmentation, improving generalization to clinical datasets. A primary limitation of our model is the tendency to hallucinate fine-grained structures when scaling to extreme resolutions (e.g., beyond $2048\times2048$), a known issue in progressive upscaling approaches where artificial detail may be introduced during denoising. Our method offers a scalable and accessible pathway for generating high-resolution medical images that can be leveraged to improve model explainability and support robustness testing. 

\begin{ack}
The authors are grateful for funding provided by the Natural Sciences and Engineering Research Council of Canada, the Canadian Institute for Advanced
Research (CIFAR) Artificial Intelligence Chairs program, Mila - Quebec AI Institute, Google Research, Calcul Quebec, and the Digital Research Alliance of
Canada.
\end{ack}

\bibliographystyle{plainnat}
\bibliography{main}

\end{document}